# Defects in Silicon Carbide as Quantum Qubits: Recent Advances in Defect Engineering

Ivana Capan

*Ruđer Bošković Institute, Bijenička 54, 10 000 Zagreb, Croatia*

**Abstract:** This review provides an overview of defects in silicon carbide (SiC) with potential applications as quantum qubits. It begins with a brief introduction to quantum qubits and existing qubit platforms, outlining the essential criteria a defect must meet to function as a viable qubit. The focus then shifts to the most promising defects in SiC, notably the silicon vacancy ($V_{Si}$) and divacancy ($V_C$-$V_{Si}$). A key challenge in utilizing these defects for quantum applications is their precise and controllable creation. Various fabrication techniques, including irradiation, ion implantation, femtosecond laser processing, and focused ion beam methods, have been explored to create these defects. Designed as a beginner-friendly resource, this review aims to support early-career experimental researchers entering the field of SiC-related quantum qubits. Providing an introduction to defect-based qubits in SiC offers valuable insights into fabrication strategies, recent progress, and the challenges that lie ahead.

1. Introduction

Quantum technologies are poised to revolutionize computing, communication, and sensing by harnessing the principles of quantum mechanics. At their foundation lie quantum bits, or qubits, which can exist in superposition states, enabling vastly enhanced computational capabilities compared to classical bits [1]. Various physical systems have been explored for qubit realization, including trapped ions, superconducting circuits, and semiconductor-based defect centers [2,3]. Among these, solid-state defects in wide-bandgap semiconductors have attracted significant attention due to their long spin coherence times, optical addressability, and compatibility with established semiconductor fabrication techniques [4]. Silicon carbide (SiC), a wide bandgap semiconductor, stands out as a promising host material due to the variety of its point defects. These point defects are commonly referred as optically active defects or color centers, and all those names will be used in this paper.

The pioneering work and understanding of color centers started with nitrogen-vacancy (NV) center in diamond [5,6,7,8]. However, color centers in SiC present notable advantages, including compatibility with wafer-scale production, precise doping control, integration with mature complementary metal-oxide semiconductor (CMOS) technology, and access to well-established nanofabrication processes. Among various color centers in SiC, the Si vacancy ($V_{Si}$) and the divacancy ($V_c$-$V_{Si}$) are the most studied [9,10,11,12,13].

A critical requirement for the successful implementation of SiC-based quantum technologies is the controlled introduction and manipulation of point defects. Various techniques have been developed to create and stabilize these defects, including ion implantation, neutron irradiation, and femtosecond laser processing [14,15,16,17,18]. Radiation-induced defects, in particular, are of great interest due to their potential for generating well-defined qubit properties through precise defect engineering. However, several challenges remain [2,4]:

i) Defect Concentration and Spatial Control: Achieving a high density of optically active defects while maintaining spatial precision is crucial for scaling quantum devices.

ii) Charge State Stability: The performance of defect-based qubits depends on stable charge states, and only specific charged states of color centers are suitable for quantum applications.

iii) High-purity material: Impurities and unintentionally introduced defects create local strain and electric field variations, leading to inhomogeneous broadening of optical and spin transition lines.

Defects in SiC have been identified as promising candidates for various quantum applications, including [9,10,11,12,16,19,20,21]:

i) Quantum Computing: Defect-based qubits can serve as scalable quantum information processors, leveraging SiC's mature fabrication ecosystem.

ii) Quantum Sensing: The spin properties of SiC color centers enable ultrasensitive detection of magnetic, electric, and thermal fields.

iii) Quantum Communication: Color centers in SiC, particularly those operating at telecom wavelengths, offer pathways for secure quantum information transfer over fiber-optic networks.

As illustrated in Figure 1, the number of research publications on quantum qubits has increased significantly over the past decade, reflecting the growing interest in defect-based quantum technologies. The data was obtained from the Web of Science Core Collection using the keyword search "quantum qubits." This trend underscores the expanding role of SiC as a key material in the development of next-generation quantum devices.

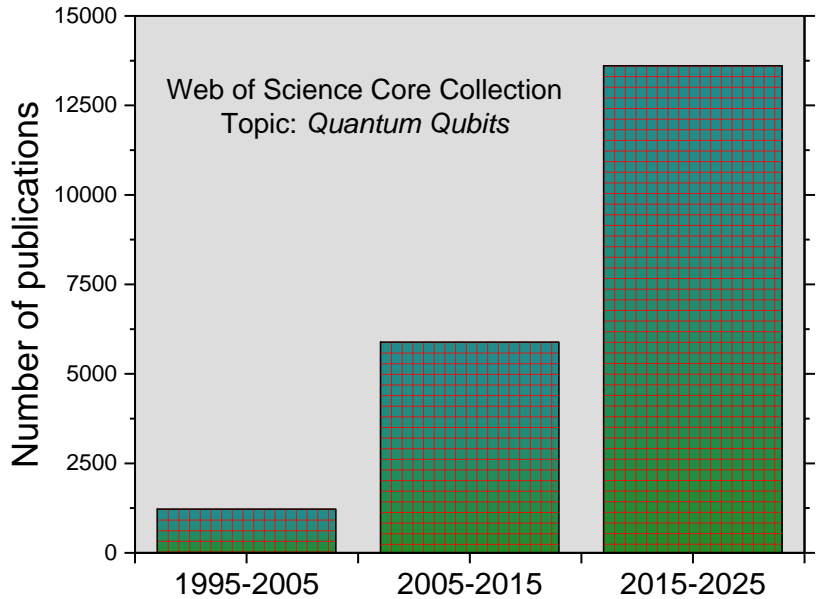

**Figure 1**. The total number of publications on quantum qubits. Data are extracted from the Web of Science Core Collection with the topic "quantum qubits".

With the rapid growth of quantum-related research, the number of review papers in this field has expanded significantly. These reviews play a crucial role in summarizing recent advancements, identifying key challenges, and outlining future research directions. Several comprehensive reviews have been published, covering various aspects of quantum defect physics, material platforms, and semiconductor qubits. Some of the most notable reviews include:

i) Zhang et al. [22], "Material Platforms for Defect Qubits and Single-Photon Emitters". This review provides a broad overview of solid-state systems pivotal for spin-based quantum technologies. It discusses advancements in solid-state spins and single-photon emitters across multiple host materials, including diamond, silicon carbide (SiC), hexagonal boron nitride (hBN), silicon, and two-dimensional semiconductors. The authors also emphasize the role of theoretical and computational methods in guiding experimental progress in defect-based quantum technologies.

ii) Chatterjee et al. [23], "Semiconductor Qubits in Practice". This paper provides a detailed analysis of semiconductor-based qubits, covering their implementation in quantum computing, quantum simulation, quantum sensing, and quantum networking. The review highlights the key challenges and progress in integrating these systems into scalable quantum architectures.

iii) Wolfowicz et al. [24], "Quantum Guidelines for Solid-State Spin Defects". This review offers a set of broad guidelines for the development and application of solid-state spin defects in quantum information processing. The authors discuss defect properties, host material characteristics, engineering opportunities, and potential pathways for improving coherence times, optical properties, and scalability.

Given the existence of these in-depth and widely cited review articles, a legitimate question arises: What unique contribution does this paper make, and how does it differentiate itself from prior work? Unlike the broader scope of the aforementioned reviews, this paper is specifically tailored for early-career experimental researchers focusing on the controllable formation of color centers in SiC for quantum qubit applications. Instead of attempting to cover the entire landscape of SiC quantum applications, this review adopts a focused and practical approach by emphasizing defect engineering techniques and their relevance to quantum qubit performance. This review provides an overview of methods for generating SiC defects, including irradiation, ion implantation, and laser processing [14,15,16,17,18]. Moreover, many existing reviews are either theory-oriented [2,4] or focused on the comprehensive material characterization that includes a whole variety of techniques such as low-temperature photoluminescence (PL), electron paramagnetic resonance (EPR) and optically detected magnetic resonance (ODMR) techniques [22,23,25,26,27,28]. This review aims to bridge the gap between theoretical studies and comprehensive material characterization by focusing on the intermediate step, defect creation.

The paper is structured as follows. Section 2 introduces the fundamental principles of quantum qubits, outlining the criteria that a defect center must satisfy to function as a viable qubit. Section 3, the core of this review, focuses on the selected SiC defects and their formation by different experimental methods. Section 4 provides a summary and conclusion. By maintaining a well-defined and specialized focus, this review serves as a valuable resource for experimentalists looking to advance their expertise in SiC-based quantum defect fabrication. It ensures that researchers can efficiently access the most relevant information to accelerate their work on next-generation quantum technologies.

## 2. Quantum qubits

This section provides an overview of the fundamental properties of quantum qubits. A quantum qubit (or simply "qubit") serves as the fundamental unit of quantum information, analogous to a classical bit. However, unlike a classical bit, which can only exist in a definite state of 0 or 1, a qubit can exist in a superposition of both states simultaneously (as graphically presented in Figure 2). These two basis states are typically represented as |0⟩ and |1⟩ [1,29]. The general state of a qubit is given by:

$$|\psi\rangle = \alpha|0\rangle + \beta|1\rangle$$

where α and β are complex probability amplitudes. The squared magnitudes of these amplitudes, $|\alpha|^2$ and $|\beta|^2$, represent the probabilities of measuring the qubit in the |0⟩ or |1⟩ state, respectively, and must always sum to one:

$$|\alpha|^2 + |\beta|^2 = 1$$

A qubit remains in its superposition state as long as it is not measured or disturbed by its environment. However, upon measurement, the superposition collapses, and the qubit assumes one of its basis states, either $|0\rangle$ or $|1\rangle$, with probabilities determined by $|\alpha|^2$ and $|\beta|^2$. This phenomenon, known as wavefunction collapse, is a fundamental aspect of quantum mechanics and plays a crucial role in quantum computation and information processing [30].

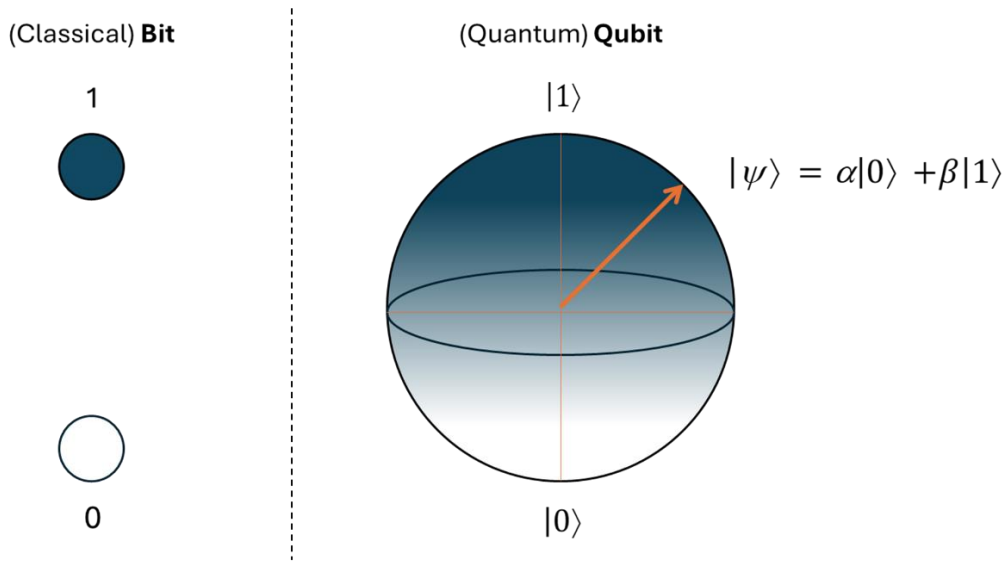

**Figure 2.** Graphical presentation of (classical) bit and (quantum) qubit.

Different physical systems are currently used as qubits. The most known are presented in Figure 3 and briefly described as follows:

i) Superconducting Qubits:

Superconducting qubits are one of the most developed and widely used platforms for quantum computing. These qubits are based on superconducting circuits that exhibit quantum behavior. Materials like niobium and aluminum are commonly used in these circuits due to their low resistance at cryogenic temperatures. Superconducting qubits offer fast operation and scalability, but they are sensitive to noise and require ultra-low temperatures for operation, typically in the millikelvin range [31, 32].

ii) Trapped Ions:

Trapped ion qubits are individual ions confined by electromagnetic fields and controlled using laser-based techniques. Commonly used materials include calcium ions ($Ca^+$), barium ions ($Ba^+$), and ytterbium ions ($Yb^+$). The internal states of these ions, such as electronic or hyperfine states, serve as qubits. Trapped ion qubits are known for their long coherence times and high-fidelity gate operations. They are ideal for quantum simulations and algorithms, but the main challenge is scalability due to the need for precise laser control and individual ion manipulation [33,34].

iii) Defect-Based Qubits:

Defect-based qubits are based on point defects in solid-state materials, where the electronic spin states of defects are used to encode information. Examples include the NV center in diamond and $V_{Si}$ or $V_C$-$V_{Si}$ centers in SiC. These defects can operate at room temperature, making them attractive for practical quantum applications. However, challenges remain in defect creation and control, as well as improving coherence times and gate fidelity [9,12,21]. For instance, NV centers in diamond are well-established for quantum sensing and communication applications, while SiC is gaining attention for scalable quantum computing [8,23].

iv) Spin-Based Qubits:

Spin-based qubits utilize the spin states of electrons or nuclei in materials like silicon and gallium arsenide (GaAs). Quantum dots, such as semiconductor quantum dots made from InAs/GaAs or Si/SiGe systems, are also commonly used. The qubit states are represented by spin-up or spin-down orientations. Spin-based qubits offer long coherence times and are compatible with existing semiconductor technology, making them attractive for scalable quantum computing. However, challenges remain in improving coupling efficiency and achieving high-fidelity gate operations [35,36].

v) Topological Qubits:

Topological qubits are based on anyonic excitations in topologically ordered materials, such as Majorana fermions in topological superconductors. Materials under investigation include superconducting nanowires made of InAs or $MoTe_2$. These qubits are robust to local noise and decoherence due to their topological nature, offering the potential for long-lived qubits. However, this platform is still experimental, and significant challenges remain in achieving reliable and scalable topological qubit systems [37,38].

Table 1 provides a summary the main advantages and challenges for all physical systems mentioned above.

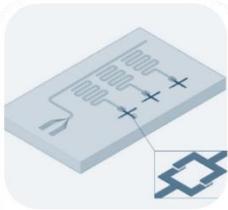
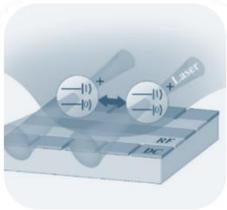
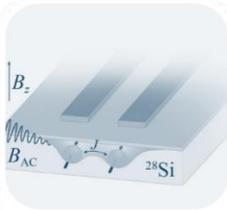
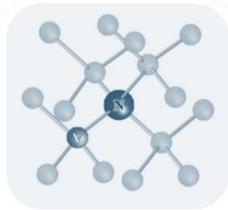
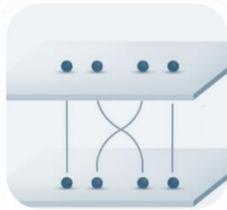

**Figure 3**. A schematic presentation of different physical systems used as quantum qubits platforms. Adapted from Ref. [39].

**Table 1.** Different quantum qubit platforms, with the main advantages and challenges [22,23,24,39].

| Type | Advantages | Challenges |
|---|---|---|
| **Superconducting qubits** | Well-established fabrication processes, strong control and readout capabilities, compatibility with microwave technology | Short coherence times (microseconds), require cryogenic cooling |
| **Trapped ions** | Long coherence times (seconds), high-fidelity gate operations | Slow gate speeds, complex laser-based control, scalability challenges |
| **Spin-based qubits (Quantum dots)** | Compatible with CMOS technology, long coherence times in isotopically purified materials | Requires precise nanofabrication, challenging control of qubit interactions |
| **Topological qubits** | Theoretically fault-tolerant, protected from decoherence | Experimentally unconfirmed, complex fabrication requirements |
| **Defect-based qubits (Solid-state qubits)** | Optical addressability, long coherence times, potential for room-temperature operation. | Difficult to create and control defects with high precision. |

It should be noted that the list of material platforms in Table 1 is not exhaustive. Other platforms, such as photonic qubits [40], are actively being investigated, and new candidates continue to emerge as research in the field advances.

With this in mind, we focus on defect-based qubits in SiC. The first fundamental question to address is: What makes a defect a viable candidate for a quantum qubit?

A key reference in answering this question is the pioneering work by Weber et al. [4] which established a well-defined set of criteria for defect-based qubits. This work outlines essential properties that both the defect and the host material must satisfy to function as a quantum qubit. A summary of these criteria is presented in Table 2.

**Table 2.** Key properties of defects and host materials for defect-based qubits [4].

| Defect properties | Host material properties |
|---|---|
| The defect must have a paramagnetic, long-lived bound state with an energy splitting suitable for manipulation via electron spin resonance | Wide bandgap: allows deep defect levels that satisfy the requirement for optical transitions. |
| The defect should have an optical cycle that allows spin initialization via spin-selective decay. | Low spin-orbit coupling: minimizes unwanted spin flips and decoherence |
| The defect's fluorescence must change depending on the qubit state, enabling efficient optical readout. | High-quality crystals: availability of bulk or thin-film single crystals to reduce imperfections and impurities. |
| Optical transitions should not be affected by interference from the host material's electronic states. | Nuclear spin-free isotopes: constituent elements should have isotopes with zero nuclear spin to minimize decoherence effects. |
| Energy separations between defect states must be large enough to prevent thermal excitations that could destroy spin coherence. | |

## 3. Defects in silicon carbide

SiC is a wide-bandgap (~3.26 eV in 4H-SiC) semiconductor with exceptional properties, making it a key material for power electronics and radiation detection applications [41,42,43,44]. It exists in more than 250 polytypes, with 4H-SiC being the most extensively studied. The 4H (hexagonal) polytype is preferred for electronic components due to its high and nearly isotropic carrier mobility, which enhances device performance [41,44]. A similar trend is observed in quantum applications, where 4H-SiC remains the dominant choice due to its favorable optical and spin properties. However, other polytypes, such as 3C-SiC (cubic) and 6H-SiC (hexagonal), are gaining increasing attention for quantum technologies, as recent studies suggest they may offer unique advantages for hosting defect-based qubits [45,46,47,48].

For a more detailed discussion of the material properties of SiC and its most commonly used polytypes (3C, 4H, and 6H), we refer readers to Ref.[41,43, 44, 49], as well as the references within. These handbooks and review papers provide comprehensive insights into the structural, electrical, and optical characteristics of SiC, making them essential reading for those interested in both fundamental and applied aspects of the material.

Among the many remarkable properties of SiC, its crystallographic structure plays a crucial role in defect engineering. Due to the presence of multiple non-equivalent lattice sites in the primitive cell, point defects in SiC can form in different configurations, significantly influencing their electronic and spin properties.

The 3C polytype, shown in figure 4a, has cubic symmetry with a single C and Si atom in the primitive cell. The 4H polytype has hexagonal symmetry and 8 atoms in the primitive cell of which 2 are non-equivalent for both Si and C (figure 4b). The 6H polytype is also of hexagonal symmetry, has 12 atoms in the primitive cell, and 3 are non-equivalent for both Si and C (figure 4c). Therefore, a single site defect in 4H has two configurations, and three in 6H-SiC. A pair defect then has four and six configurations in 4H and 6H-SiC, respectively. The non-equivalent sites are denoted as *h* (hexagonal) and *k* (quasi-cubic) [50,51].

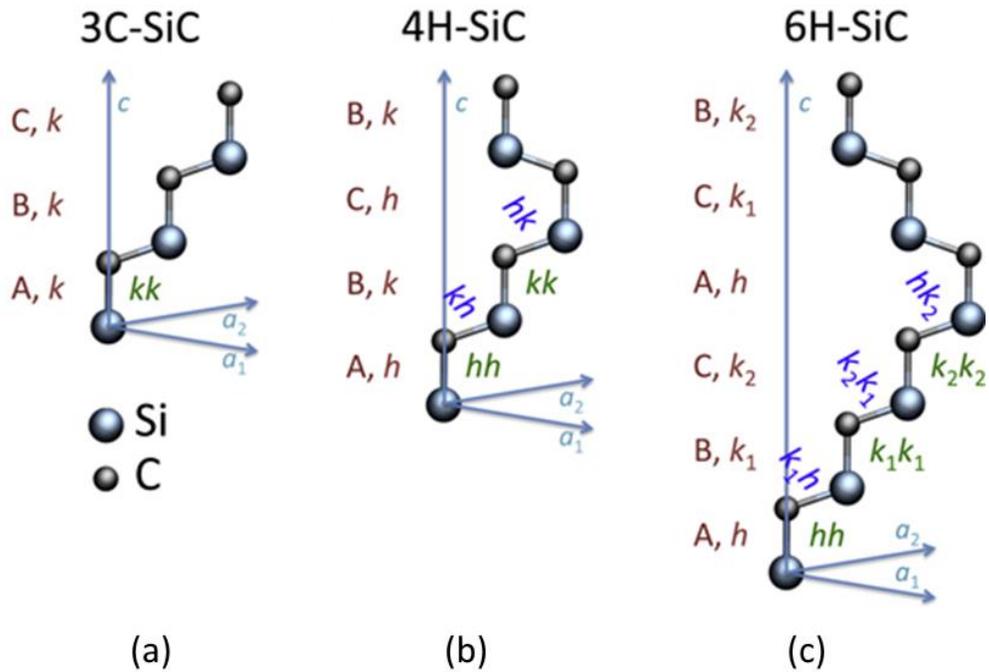

**Figure 4.** The primitive cells of 3C, 4H and 6H-SiC. Red uppercase letters indicate the stacking of Si–C double layers, while lowercase letters denote whether the double layers and their immediate surroundings follow a cubic (*k*) or hexagonal (*h*) stacking order. Green lowercase letter pairs represent variants of a pair defect. Figure taken from Ref. [51].

Figure 5 shows the structure of $V_{Si}$ and $V_C$-$V_{Si}$ in 4H-SiC. Divacancy has four, denoted as *hh*, *kk*, *hk* and *kh* while silicon vacancy has two non-equivalent sites denoted as *k* and *h*.

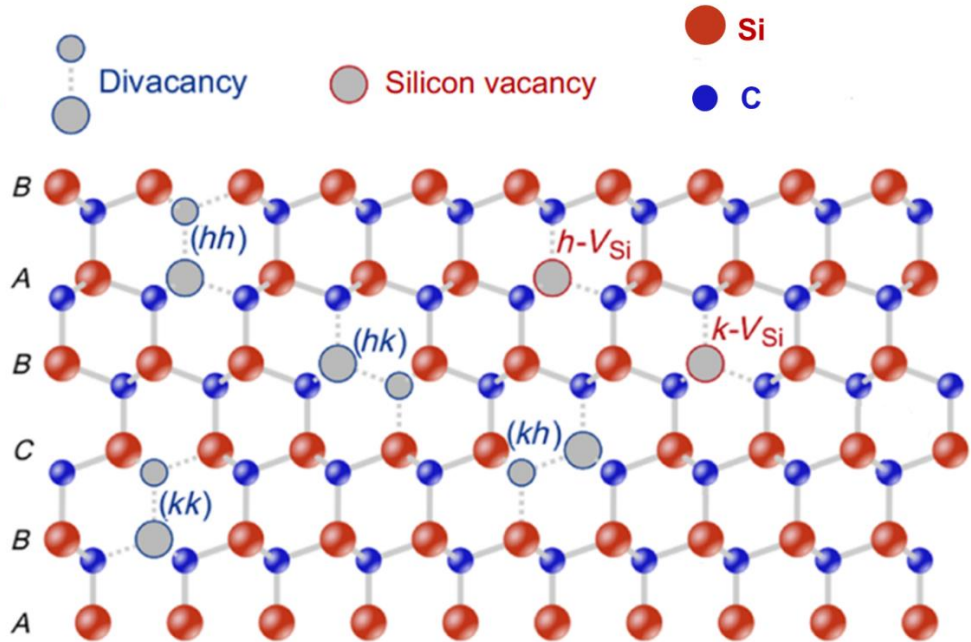

**Figure 5.** The 4H-SiC lattice structure showing the non-equivalent configurations of the divacancy ($V_C$-$V_{Si}$) and, silicon vacancy ($V_{Si}$). Black uppercase letters indicate the stacking of Si–C layers (red and blue spheres represent Si and C atoms, respectively). Adapted from Ref. [52].

Adding to the complexity, each defect can exist in multiple charge states within the SiC bandgap, further influencing their electronic transitions, optical emission properties, and spin characteristics. The ability to precisely control and identify these charge states is critical for optimizing defect-based qubits, as unwanted charge-state fluctuations can degrade coherence and readout fidelity [53, 54]. This remains one of the most significant challenges: the selection and stabilization of the desired charge states. As discussed in the following sections, color centers in SiC can exist in multiple charge states, but only specific ones are suitable for quantum applications.

The first step in the creation of defect-based qubits is the controlled introduction of defects into SiC. This is typically achieved through ion implantation or high-energy irradiation, both of which introduce lattice vacancies and interstitials that can later form stable defect centers. Figure 6 (left side) schematically illustrates ion implantation and high-energy irradiation, two widely used techniques for defect engineering in SiC. In the case of ion implantation, a high-energy ion beam (such as He, C, N, or other species) is used to displace Si and C atoms in a specific volume of the SiC material. The implantation energy determines how deep the defects are introduced, enabling precise spatial control over defect placement. However, a significant drawback of this method is that other unwanted defects are created along the ion cascade, which can impact the material's properties. Unlike ion implantation, which introduces defects in a localized manner, high-energy irradiation (e.g., electron or neutron irradiation) generates vacancies and interstitials uniformly throughout the material. This technique ensures a more homogeneous distribution of defects, but it lacks the spatial precision offered by ion implantation.

To determine implantation or irradiation conditions, such as estimating the depth profile and distribution of vacancy-related defects in SiC, various simulation tools are commonly employed. For ion implantation,

SRIM code [55] is typically used, while FLUKA code [56] is often applied for simulating neutron-induced damage. A representative example is provided in the study by Brodar et al. [57], where they compared the spatial distribution of vacancy-related defects introduced by ion implantation (He$^+$, C$^+$) and neutron irradiation within the volume of 4H-SiC.

However, in both cases (ion implantation and irradiation), a high-temperature annealing step is often required to rearrange displaced atoms, remove unwanted defects, and stabilize the desired charge state of the defect centers.

Figure 6 (right) schematically illustrates the most advanced approach to defect engineering, where defects are introduced at precise locations and with controlled concentrations. Achieving such a level of precision is crucial for the development of scalable quantum technologies but remains a considerable challenge. Among the most promising techniques moving toward this goal are laser writing and focused ion beam (FIB) implantation.

Laser writing utilizes ultrafast femtosecond laser pulses to generate point defects at targeted positions within the SiC lattice and has demonstrated high spatial precision in defect creation [14]. In contrast to conventional broad-beam ion implantation, FIB implantation allows for the delivery of individual ions into specific locations, enabling precise control over defect placement with minimal collateral damage [17,18]. Despite its advantages in spatial accuracy, FIB implantation typically requires post-implantation annealing to activate the desired defect states and repair lattice damage.

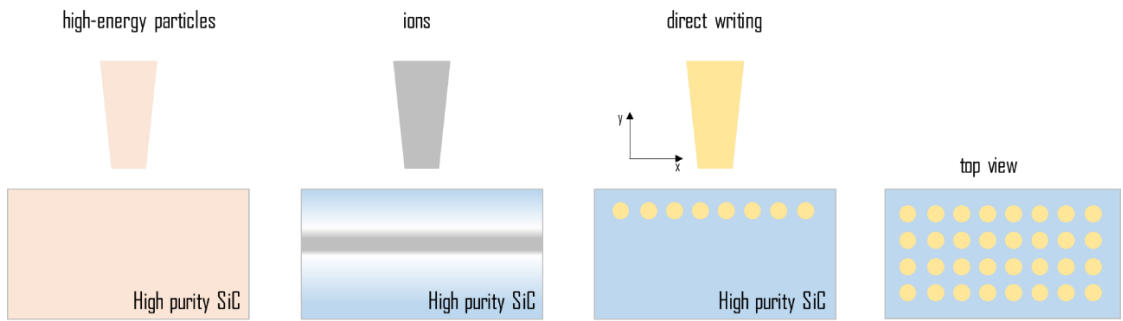

**Figure 6.** Schematic illustration of experimental methods used to introduce defects (from left to right): high-energy particle irradiation, ion implantation, and direct writing techniques such as laser writing or focused ion beam. On the far right, a top-view image of the sample after direct writing or FIB implantation, reveals a well-defined array of defects created with high spatial precision.

### 3.1 Silicon Vacancies ($V_{Si}$)

The negatively charged $V_{Si}$ is one of the fundamental intrinsic defects in SiC. $V_{Si}$ is an optically active defect and plays a crucial role in quantum applications. Here, an overview of key experimental methods used for the creation of silicon vacancies in SiC is given. The commonly employed methods include irradiation and ion implantation, both of which typically require post-treatment annealing to remove undesired defects introduced during the process and to reach the preferable state. Across various available studies, 4H-SiC has been selected as the preferred SiC polytype.

Morioka et al. [16] irradiated 4H-SiC epitaxial layer with 2 MeV electrons to generate isolated $V_{Si}$ centers. To ensure that vacancies were in the desired negatively charged state, a slightly n-type doped material was used. The samples were then annealed at 300°C for 30 min to eliminate unwanted defects. Fuchs et al. [58] also used 4H-SiC, but instead of electron irradiation, they employed reactor neutrons with energies ranging from 0.18 to 2.5 MeV. The highest neutron irradiation fluence, $5 \times 10^{17}$ cm$^{-2}$, required a multi-step thermal annealing process at temperatures ranging from 125 to 700°C for 90 min [58].

Pavunny et al. [18] have performed room-temperature 100 keV Li ion beam implantation in the form of 10 × 10 square matrix arrays of spots. Figure 7 shows PL spectrum at 4.8 K acquired from an individual spot of a 10 × 10 array implanted at a dose of $1.2 \times 10^{14}$ ions/cm$^2$. The single spot (600 ions/spot) is identified (red circle) in the low-temperature confocal fluorescence map (Fig.7, inset).

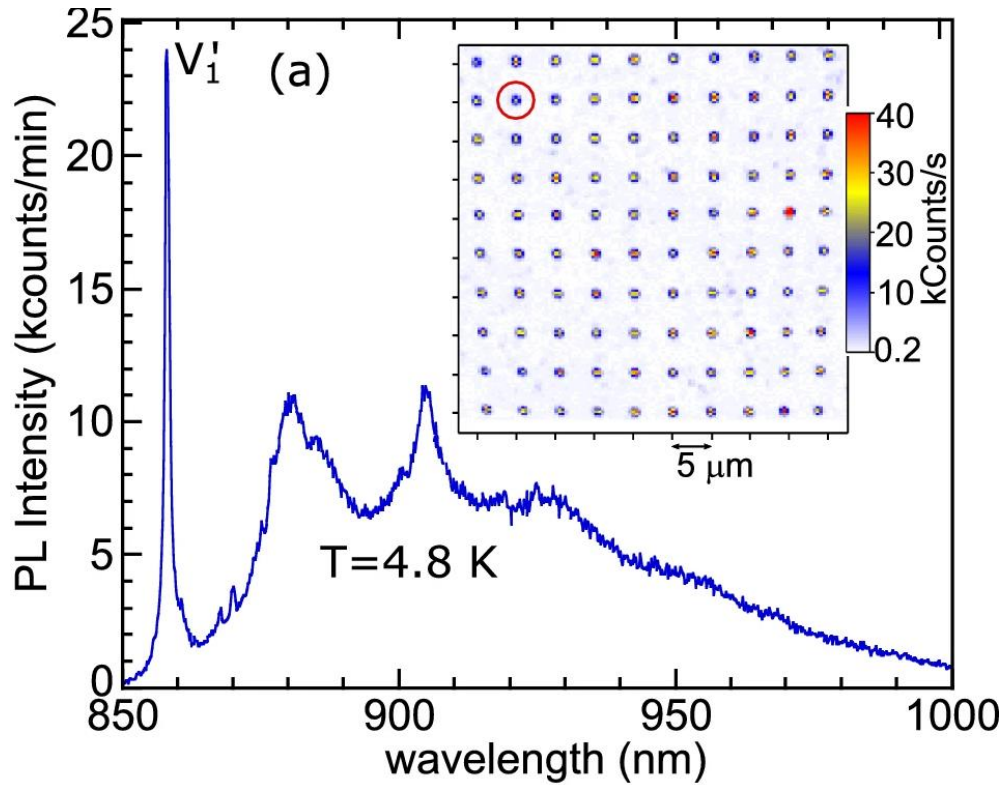

**Figure 7.** PL spectrum at 4.8 K acquired from an individual spot of a 10 × 10 array implanted at a dose of $1.2 \times 10^{14}$ ions/cm$^2$ (600 ions/spot) which is identified (red circle) in the low temperature confocal fluorescence map (inset). Figure taken from Ref. [18]

Kraus et al. [17] demonstrated a method by using focused proton beams for controlled defect generation without the need for pre- or post-annealing treatments. By matching the proton energy to the stopping power of the material, the generation depth and resolution could be precisely controlled, and the number of defects in a specific volume was tunable. Similar results were achieved by laser writing [14]. The focused ion beam and laser writing methods represent the most promising methods for highly localized and well-controlled defect formation in SiC. The key experimental methods used for introducing $V_{Si}$ into SiC are summarized in Table 3.

**Table 3**. Experimental methods used for single $V_{Si}$ creation. Data adapted from Ref. [25].

| Method | Source | Energy | Reference |
|---|---|---|---|
| Electron Irradiation | $e^-$ | 2 MeV | [16] |
| Neutron irradiation (reactor) | $n^0$ | 0.18–2.5 MeV | [58] |
| Ion implantation | $He^+$ | 6 keV | [59] |
| | $He^+, C^+$ | 20 keV | [26] |
| Focused ion beam | $Li^+$ | 80 keV | [25] |
| | $Li^+$ | 100 keV | [18] |
| | $He^+$ | 30 keV | [60] |
| Focused proton beam | $H^+$ | 2 MeV | [17,61] |
| Laser writing | -- | -- | [14,62] |

Silicon vacancy in 4H-SiC gives rise to well-known PL lines, commonly referred to as V1 (*h* site) and V2 (*k* site) [50]. Moreover, V1 has a high temperature companion labelled as V1' [50], as shown in Figure 7. This leads us to another noteworthy issue that adds to the complexity of defect engineering. The high-energy particle irradiation and ion implantation also introduce electrically active defects into SiC [15,43,44,64,65]. Among them is the silicon vacancy, which introduces two deep levels into the bandgap. Those levels are known as S1 and S2 and they have been assigned to different charge states of $V_{Si}$ at *h* and *k* site, (-3/=) and (=/0), respectively [15,44]. Bathen et al. [15] have presented a very rare if not unique study where optically and electrically active defects (i.e. silicon vacancies) are compared, offering valuable insight into their fundamental properties and potential for quantum technology applications [15]. In their study, they successfully correlated the optical V1/V1' signals and the electrical S1/S2 signals.

To summarize, once introduced in SiC, $V_{Si}$ has been comprehensively characterized using PL, EPR and OMDR [10,11,12,13,14,15,16,17,18,19,27,28,50,51,66]. Readers are strongly encouraged to explore the referenced papers for further insights. Ongoing research continues to focus on optimizing spatial precision, charge state control, and process efficiency to meet the growing demands of quantum applications. It is important to note that, for the sake of clarity, numerous other equally valuable studies have not been included in this summary.

3.2. Divacancies ($V_{Si}$-$V_c$)

Together with negatively charged $V_{Si}$, a neutral divacancy $V_{Si}$-$V_c$ is considered as a very promising candidate for quantum applications. A divacancy is formed when two neighboring silicon and carbon atoms are missing from the SiC lattice as shown in Figure 5.

Like V$_{si}$, divacancies are often introduced through high-energy irradiation or ion implantation. Shafizadeh et al. [67] have used high-purity 4H and 6H SiC samples and studied the selection rules of divacancies which were introduced by electron irradiation and subsequent annealing. Sun et al. [68] have used proton irradiation 250 keV to introduce both, silicon vacancies and divacancies. Almutairi et al. [69] have achieved direct writing of divacancy arrays in 4H-SiC using the femtosecond laser and subsequent thermal annealing. He et al. [28] fabricated single divacancy arrays in 4H-SiC using the He FIB with an energy of 30 keV. The samples were annealed after implantation at 500 ◦C for 2 h to generate monovacancies, followed by a post-annealing at 900 ◦C for 1 h to facilitate monovacancy migration and divacancy formation. Additionally, the authors have used C ion with the same energy, but enhanced optical and spin characteristics were notably observed after He ion compared to C ion implantation. Figure 8 shows confocal scanning image of the introduced divacancy arrays.

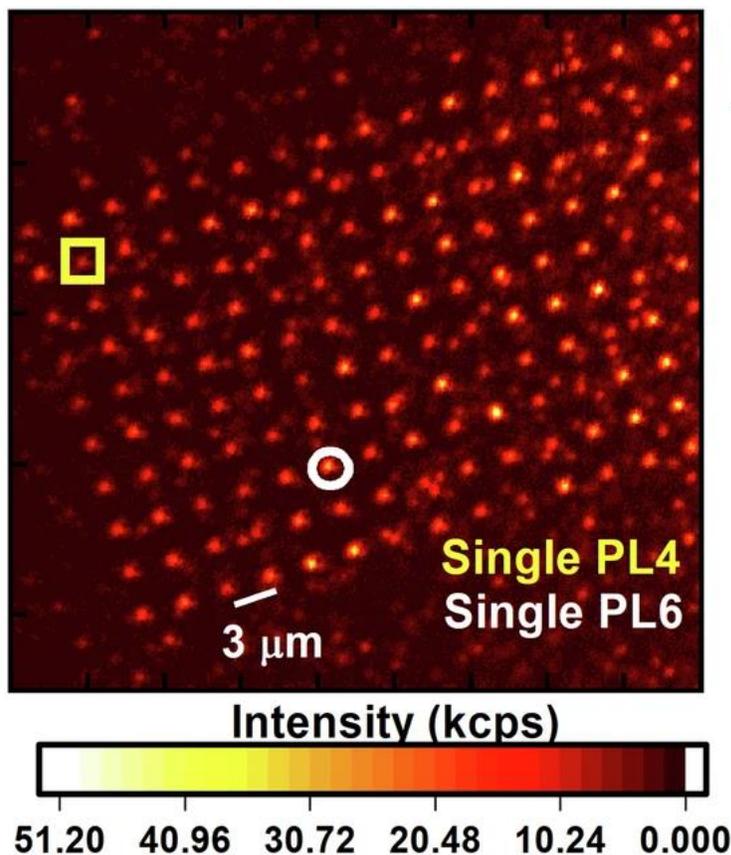

**Figure 8.** Confocal scanning image displaying the divacancy arrays generated using a dose of 300 ions/spot. The white bar represents a distance of 3 μm. The yellow square and white circle enclose the individual PL4 and PL6, respectively. Figure taken from [28]

As already said, the divacancy has four configurations in 4H-SiC (*hh*,*kk*,*hk* and *kh*) which are shown in Figure 5. Correspondingly, there are four PL lines in 4H-SiC known as Pl1, PL2, PL3, and PL4. [22,54,67]. The PL5 and PL6 lines, which have been previously detected but not convincingly assigned, are suggested to be related to the modified divacancies located at or near Frank-type stacking faults [28]. Figure 9 shows PL spectrum of the various divacancies in 4H-SiC with six PL lines being observed. While investigating the individual divacancy arrays in 4H-SiC introduced by He FIB, He et al. [28] have studied the so-called modified divacancies, PL5 and PL6. Their results have indicated PL6 as a defect with exceptional properties for quantum applications.

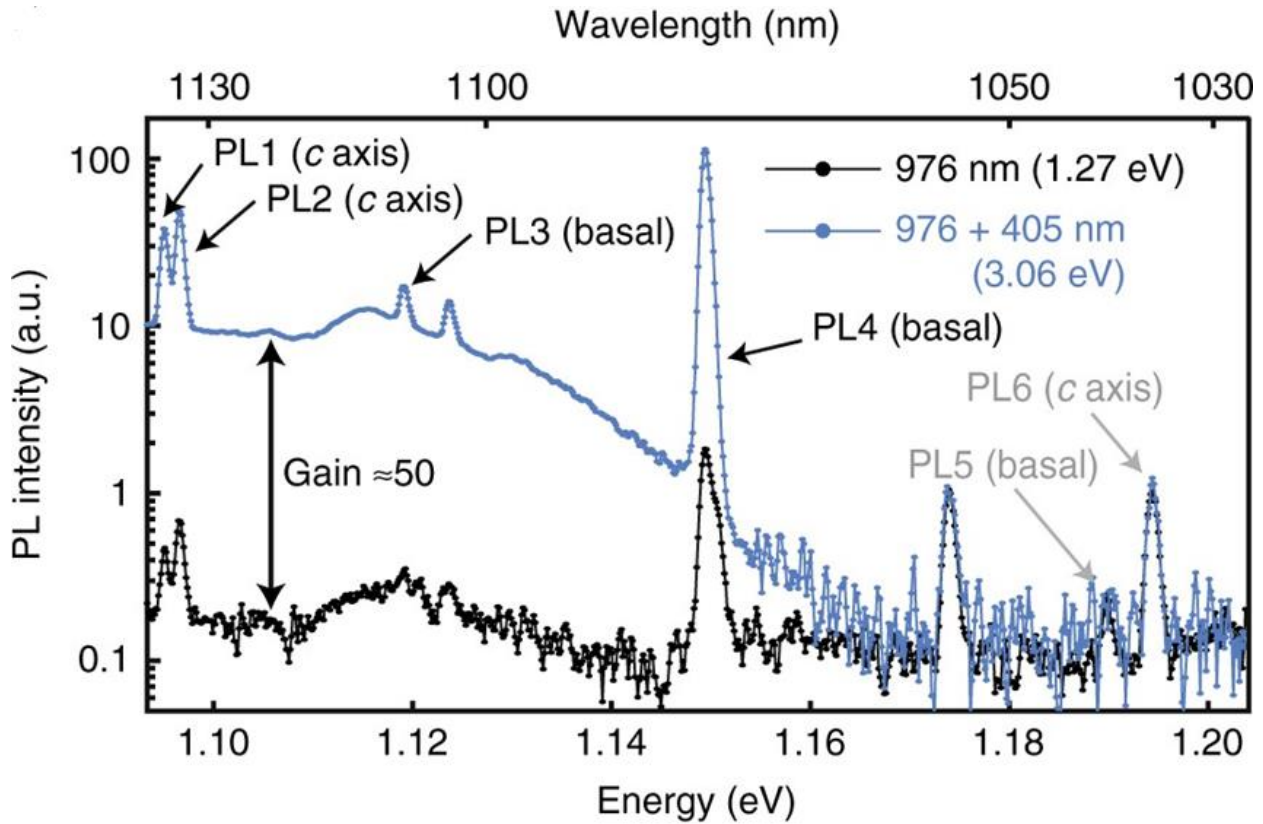

**Figure 9.** PL spectrum with 976 nm excitation of the various divacancies in 4H-SiC. Figure taken from Ref. [70]. Six PL lines are detected.

Significant progress in divacancy defect engineering has been achieved by Wolfowicz et al. [70]. They have achieved the optical charge state control of divacancy in 4H-SiC. Using the above-bandgap excitation they have efficiently converted divacancy toward its neutral-charge state, and more importantly drastically increased the PL intensity, without affecting the spin properties.

Another well-studied defect in SiC is the negatively charged nitrogen-vacancy (NV). This defect consists of a nitrogen shallow donor at a carbon lattice site paired with a neighboring silicon vacancy (Figure 10).

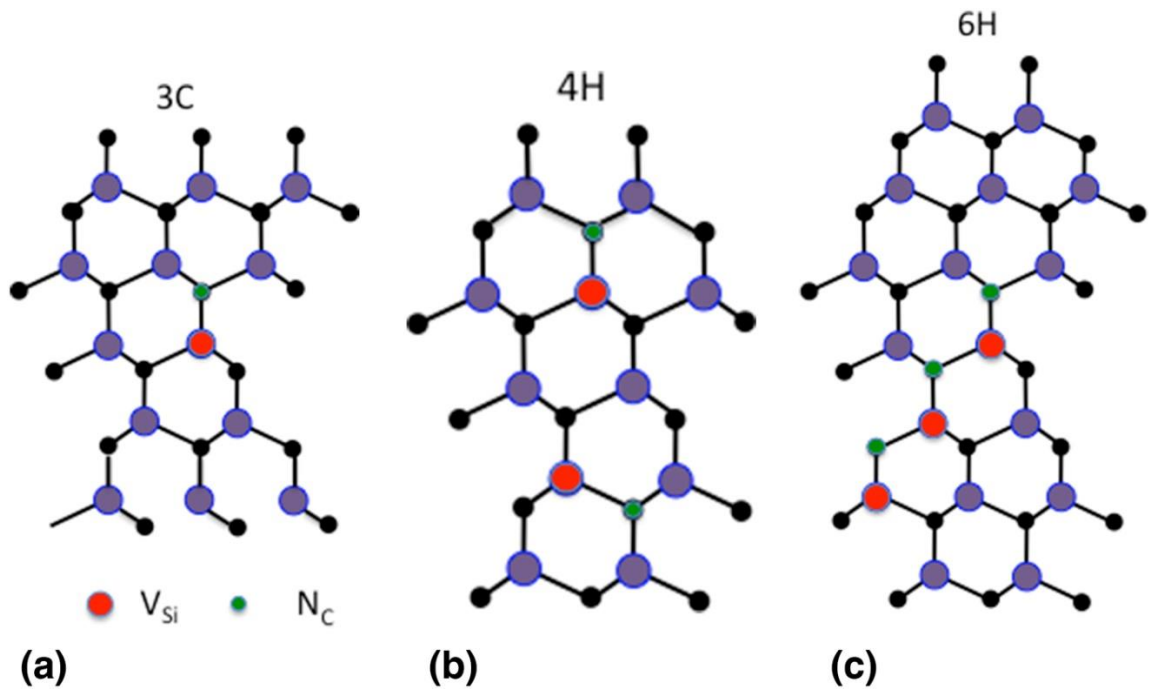

**Figure 10**. (a) Microscopic structure of the NV center in 3C-SiC. (b) Axial and basal NV centers in 4H-SiC. (c) The three distinct axial NV centers in 6H-SiC. (purple and black spheres represent Si and C atoms, respectively). Figure taken from Ref. [71].

Wang et al. [72] employed nitrogen ion implantation with an incident energy of 30 keV to introduce NV centers in 4H-SiC. To optimize the formation of divacancies, they systematically varied the annealing temperature from 800°C to 1050°C. Their study successfully demonstrated coherent control of NV center spins and characterized the fluorescence and ODMR properties of single NV centers in 4H-SiC at room temperature. By refining the annealing conditions, they achieved a significant increase in NV center concentration. Jian et al. [73] utilized high-energy (12 MeV) proton irradiation to generate NV centers in 4H-SiC. The irradiated samples were subsequently annealed at elevated temperatures, leading to the formation of NV centers with an estimated concentration of approximately $10^{16}$cm$^{-3}$. Bardeleben et al. [74] investigated the creation of NV centers across multiple SiC polytypes (3C, 4H, and 6H-SiC) using high-energy electron irradiation followed by post-irradiation annealing. Their study provided valuable insights into the defect formation processes and the impact of different irradiation methods on NV center generation in SiC.

The summary of the main experimental methods used for creating divacancies and NV centers in SiC is given in Table 4.

**Table 4.** Experimental methods used for creation of divacancies and NV centers in SiC.

| Defect | Method | Source | Energy | Reference |
|---|---|---|---|---|
| **VV** | Electron irradiation | $e^-$ | 2 MeV | [67] |
| **VV** | Femtosecund laser writing | -- | -- | [68] |
| **VV** | Focused ion beam | $He^+$ | 30 keV | [28] |
| **VV** | Ion implantation | $H^+$ | 250 keV | [69] |
| **NV** | Electron irradiation | $e^-$ | >MeV | [74] |
| **NV** | Proton irradiation | $H^+$ | 12 MeV | [73] |
| **NV** | Ion implantation | $N^+$ | 30 keV | [72] |

While the aforementioned defects have demonstrated the greatest potential for quantum applications, new defect candidates are continuously emerging. Among these are the carbon antisite-vacancy pair (CAV) and chromium (Cr)-related defects in SiC. CAV defects are typically introduced via irradiation [75,76], while Cr-related centers can be incorporated either through Cr ion implantation [77]. Although the fabrication techniques and charge state control for these emerging defects are still underdeveloped, continued advancements in the field suggest that substantial improvements can be expected in the coming years.

4. Conclusions

The remarkable potential of silicon carbide (SiC) and its point defects for quantum applications has been convincingly demonstrated over the past decade. Notable advancements have been made in defect engineering, particularly in the precise and reproducible creation of color centers such as silicon vacancy and divacancy. Techniques such as focused ion beam (FIB) implantation and femtosecond laser writing have emerged as powerful tools. These methods offer enhanced control over the spatial localization and density of defects, which is essential for scalable quantum device fabrication. Additionally, a significant milestone in defect engineering has been the achievement of optical charge state control in divacancy centers, marking a critical step toward reliable quantum operation. The level of understanding and control achieved for defects such as silicon vacancy and divacancy has not yet been reached for other, newly emerging defects. However, these emerging defects may still find unique applications within the broader field of quantum technologies, even if not directly as qubits

Among the different polytypes, 4H-SiC continues to lead the field due to the availability of high-purity material and well-established fabrication protocols. Nonetheless, the potential of other polytypes, such as 3C-SiC and 6H-SiC, remains largely untapped. These materials present unique properties and opportunities that warrant further investigation, as they may offer alternative pathways for the development of scalable quantum technologies.